\documentclass[10pt,twocolumn,letterpaper]{article}

\usepackage{wacv}
\usepackage{times}
\usepackage{epsfig}
\usepackage{graphicx}
\usepackage{amsmath}
\usepackage{amssymb}

\usepackage{multirow}
\usepackage{multicol}
\usepackage{caption}
\usepackage{nicefrac}
\usepackage{subfig}
\usepackage{makecell}
\usepackage{color, colortbl}
\usepackage[table]{xcolor}
\definecolor{lightgray}{rgb}{0, 0, 0}

\usepackage{amsmath,amsfonts,bm}

\def\eqref#1{equation~\ref{#1}}

\def\1{\bm{1}}

\def\rvu{{\mathbf{i}}}

\def\rvu{{\mathbf{u}}}

\def\rvx{{\mathbf{x}}}
\def\rvy{{\mathbf{y}}}
\def\rvz{{\mathbf{z}}}

\DeclareMathAlphabet{\mathsfit}{\encodingdefault}{\sfdefault}{m}{sl}
\SetMathAlphabet{\mathsfit}{bold}{\encodingdefault}{\sfdefault}{bx}{n}

\usepackage{sidecap}

\usepackage[nolist,nohyperlinks]{acronym}
\acrodef{MDF}[MDF]{{multi-scale discriminative features}}
\acrodef{SM}{supplementary materials}
\acrodef{SISR}{single-image super resolution}
\acrodef{HR}{high-resolution}
\acrodef{LR}{low-resolution}
\acrodef{JND}{just-noticeable-difference}
\acrodef{CNN}{Convolutional Neural Networks}
\acrodef{MSE}{mean-squared error}
\acrodef{GAN}{generative adversarial network}

\usepackage[pagebackref=true,breaklinks=true,letterpaper=true,colorlinks,bookmarks=false]{hyperref}

\newcommand{\red}[1]{{\textcolor{red}{#1}}}
\newcommand{\blue}[1]{{\textcolor{blue}{#1}}}

\def\wacvPaperID{784} 

\wacvfinalcopy 

\ifwacvfinal\fi

\ifwacvfinal
\usepackage[breaklinks=true,bookmarks=false]{hyperref}
\else
\usepackage[pagebackref=true,breaklinks=true,colorlinks,bookmarks=false]{hyperref}
\fi
\usepackage[accsupp]{axessibility}

\ifwacvfinal\fi
\begin{document}

\title{Training a Task-Specific Image Reconstruction Loss}
\author{Aamir Mustafa \qquad Aliaksei Mikhailiuk \qquad Dan Andrei Iliescu \\
Varun Babbar \qquad Rafa\l{} K. Mantiuk\\
University of Cambridge, UK\\
\tt \small{Project website: \url{https://www.cl.cam.ac.uk/research/rainbow/projects/mdf/}}
}

\maketitle
\ifwacvfinal\thispagestyle{empty}\fi
\begin{abstract}

The choice of a loss function is an important factor when training neural networks for image restoration problems, such as single image super resolution. The loss function should encourage natural and perceptually pleasing results. A popular choice for a loss is a pre-trained network, such as VGG, which is used as a feature extractor for computing the difference between restored and reference images. However, such an approach has multiple drawbacks: it is computationally expensive, requires regularization and hyper-parameter tuning, and involves a large network trained on an unrelated task. Furthermore, it has been observed that there is no single loss function that works best across all applications and across different datasets. In this work, we instead propose to train a set of loss functions that are application specific in nature.
Our loss function comprises a series of discriminators that are trained to detect and penalize the presence of application-specific artifacts. 
We show that a single natural image and corresponding distortions are sufficient to train our feature extractor that outperforms state-of-the-art loss functions in applications like single image super resolution, denoising, and JPEG artifact removal. Finally, we conclude that an effective loss function does not have to be a good predictor of perceived image quality, but instead needs to be specialized in identifying the distortions for a given restoration method. 
\end{abstract}

\vspace{-0.5cm}
\section{Introduction}


The success of deep learning over the past several years has led to extensive use of Convolutional Neural Networks (CNNs) on a wide range of image restoration tasks, such as single-image super resolution or denoising. One of the critical choices effecting CNNs performance is the loss function. A popular mean-squared error (MSE or $L_2$) loss often results in blurry, splotchy \cite{Zhao2017} or unnatural looking images as the reconstructed image tends to be an average of potential solutions, which may not lie on the natural image manifold \cite{Blau2018CVPR}. Generative Adversarial Networks (GANs) \cite{Goodfellow2014} can ensure that resulting images lie on such a manifold, but when used alone, may result in images that are substantially different from the input \cite{Blau2018CVPR}. Furthermore, GANs are challenging to train due to the instability of their optimization problem. 

A new category of loss functions, which has recently gained noticeable popularity, employs neural networks as feature extractors. Most commonly, the loss is computed as the $L_2$ distance between the activations of the hidden layers of a trained image classification network (e.g. a VGG network \cite{Simonyan2015}). 
Such losses have been successful in training learning-based image restoration models. However, a major drawback of these loss functions is that they use large image classification networks as feature extractors. This not only makes the training process memory intensive, but also focuses on image regions which are more salient for the task of image classification. Recently, Zhang \textit{et al.} \cite{Zhang2018} tried to overcome this shortcoming by introducing a Learned Perceptual Image Patch Similarity (LPIPS) metric. They calibrated existing pre-trained classification networks on a new dataset of human perceptual similarity judgments. However, this approach still requires an extensive dataset for training the feature extractor. 
Furthermore, LPIPS/VGG features need to be complemented with an $L_2$ loss term to offer acceptable performance, which involves the need to carefully tune weights of both loss terms.

In this work, we explore the question of what makes a good loss function for an image restoration task, such as single image super resolution, denoising and JPEG artifact removal. It has been observed that there is no single loss function that works best across different applications \cite{delbracio2020projected,Ding2021,jo2020investigating}. This motivates the need to a novel set of loss functions that do not aim to be universal, but instead are task-specific. In this work, we propose our task-specific Multi-Scale Discriminative Feature (MDF) loss function, which is trained on a single natural image. Despite very lightweight training, our loss outperforms popular feature-wise (perceptual) losses, which have been trained on very large datasets. This is possible, because our loss does not learn the distribution of natural images but instead is trained to penalize the task specific distortions at different image scales. The latter task is more relevant for a loss function and much easier, thereby can be learned with as little as a single training image. Furthermore, we show that our loss function performs better as a regularization term in an adversarial setting than the VGG loss.
An extensive comparison in terms of objective metrics and subjective image quality study shows that our loss function outperforms the state-of-the-art losses for varied image restoration tasks across different datasets.

\section{Related work}


In recent years, the search of an optimal perceptual loss function has gained much attention. Below, we differentiate between hand-crafted losses, which rely on existing metrics, feature-wise losses, where image statistics are extracted using deep learning models, and distribution losses, where the loss pushes the solution to the manifold of natural images.

\noindent \textbf{Hand-crafted losses:} 
Zhao \textit{et. al} \cite{Zhao2017} have studied visual quality of images produced by the image super-resolution, denoising and demosaicing algorithms using $L_2$, $L_1$, SSIM \cite{Wang2004} and MS-SSIM \cite{Wang2003} as loss functions. Images, produced by the algorithms trained with the combination of $L_1$ and MS-SSIM losses attained the best quality as measured by objective quality metrics. That result was closely followed by the $L_1$ loss used on its own. Ding \textit{et al.} \cite{Ding2021} compared a number of image quality metrics used as a loss function in image reconstruction methods. They found that many of the popular quality metrics do not have properties that could warrant good reconstruction results. 

\noindent \textbf{Feature-wise losses:}
Similarity between the reference and the generated image can be computed in the feature space of deep CNNs. This class of losses are often called \emph{perceptual losses} as they are meant to optimize the perceptual quality rather than the pixel differences. However, since these loss functions do not explicitly model perceptual processing, we use a more descriptive name of feature-wise losses.

Authors in \cite{johnson2016perceptual} used the $L_2$ norm between the features of the reference and test images extracted from the VGG \cite{Simonyan2015} network as a loss function to train style-transfer and super-resolution algorithms. Here the VGG network was trained on ImageNet dataset \cite{Russakovsky2015}. Authors in \cite{Zhang2018} (LPIPS) have noted that features learned while training the network for image quality assessment task might better capture perceptual similarity between the target and generated image. The work used the features of several networks (untrained VGG, VGG trained on the ImageNet dataset, and on image quality dataset) to predict image quality. The authors observed that hidden representations of all tested deep models encode features important for perceptual similarity. However, deep features at various levels vary in their capacity to model perceived quality. The work of \cite{Tariq2020} proposed a methodology for selecting deep features of pre-trained CNNs that have the strongest relationship with the perceptual similarity.

However, training image restoration algorithms reliant only on the features extracted from the deep network as a loss is unstable \cite{Blau2018CVPR}. Due to pooling in the hidden layers, the network implementing the function is often not bijective, meaning that different inputs to the function may result in identical latent representations \cite{Blau2018CVPR}. Therefore, feature-wise losses are often used in conjunction with a regularization term, such as $L_2$ or $L_1$ norms, and require careful tuning of the weights of each loss component. Delbracio \textit{et. al} \cite{delbracio2020projected} proposed a modification to penalize the VGG features of the reference and the predicted image based on the 1D-Wassertein distance \cite{villani2009optimal,peyre2019computational}. However, this method again relies on $L_1$ normalization for training to achieve acceptable results.

\noindent \textbf{Distribution loss (GAN):}
Many image restoration algorithms are inherently ill-posed. For example images produced by super-resolution or denoising algorithms can have acceptable perceptual quality while not precisely matching the ground-truth. These algorithms can be optimized to produce images that lie on the natural image manifold, constrained by the similarity to the ground truth distribution. To ensure that the first requirement is met, many works have relied on GANs \cite{Goodfellow2014}. In such a setting, the image-generation algorithm has two loss terms: the discriminator, trained to differentiate between the generated and natural images, and a term constraining the generator network to produce images close to the ground truth. In \cite{yeh2016semantic,Isola_2017_CVPR} authors used $L_1$ norm to regularize the training. Similarly the works of  \cite{Dosovitskiy2016,ledig2017photo} used the feature-wise VGG-based loss to constraint the generator. Some other works combined both hand-crafted and feature-wise losses \cite{Sajjadi2017,Wang_2018_ECCV_Workshops}. Others have introduced regularization based on the feature loss of the discriminator \cite{tej2020enhancing,jo2020investigating}. To avoid regularization in training for SISR the work of \cite{Bahat2020} proposed to use the consistency enforcing module. The module can wrap any SISR architecture, making it satisfy the consistency constraint -- a down-sampled version of the image reconstructed with the network must be close to the low-resolution input.

Inability of the losses to generalize over different image restoration applications and over varied datasets raises the need of a task-specific loss function. In the following section, we introduce our proposed loss function which is trained to identify and thereby successfully remove the distortions for a given image restoration task in hand.

\begin{figure*}[t]
    \centering
    \includegraphics[width=.9\textwidth]{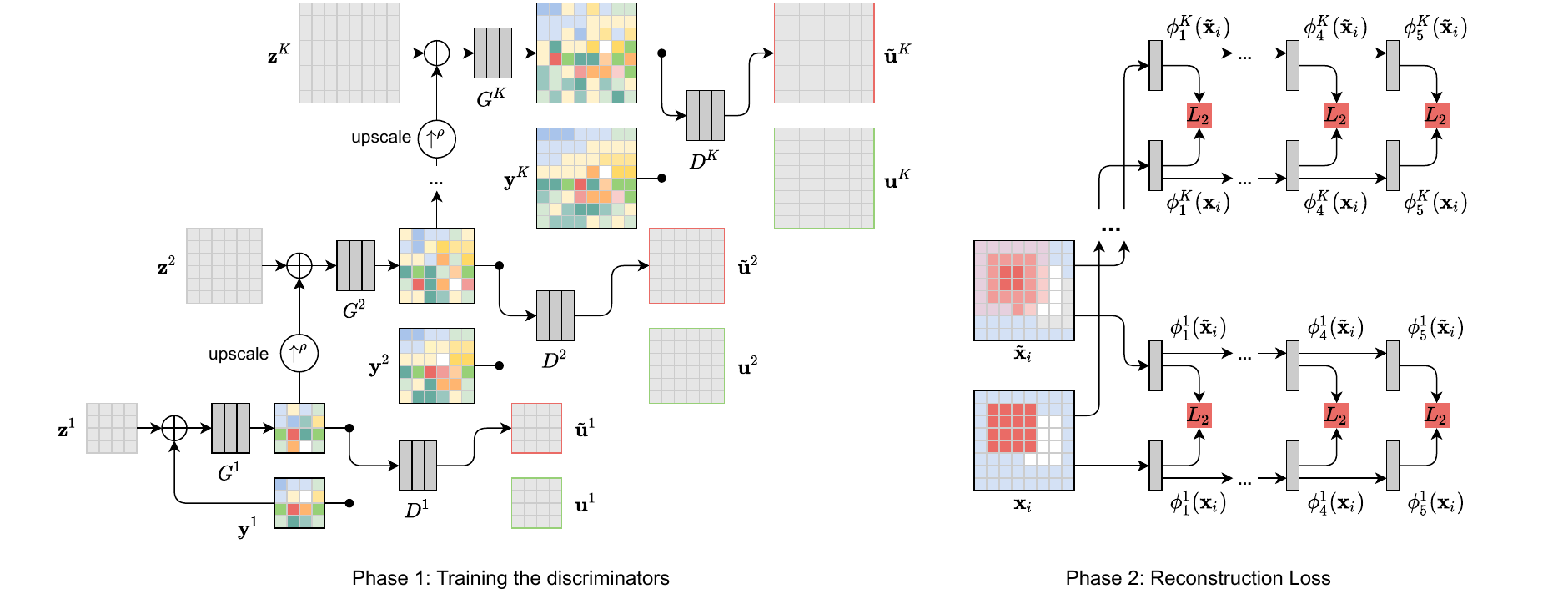}
    \vspace{-0.3cm}
    \caption{Graphical illustration of the two phases of our loss. \textbf{Phase 1} denotes the adversarial training of the discriminators. The generated image is produced by the scale-specific generator $G^k$, which takes as input the upscaled output of the previous level $\tilde{\rvy}^{k-1} \uparrow^\rho$ added with the task specific distortions $\rvz^k$. For SISR, no distortions are added ($\rvz^k = 0$). The levels are sequentially trained from the coarsest scale to the finest. In \textbf{Phase 2}, the discriminators are frozen and used as feature extractors over whose outputs an $L_2$ distance is measured between the ground-truth training image $\rvx_i$ and the restoration output $\tilde{\rvx}_i$. The distance is measured between the two images at every scale $k$ and intermediate layers of the discriminator $||\phi^k_l(\rvx_i) - \phi^k_l(\tilde{\rvx}_i)||^2_2$. 
    }
    \label{fig:main_diag}
    \vspace{-0.35cm}
\end{figure*}

\section{Multi-Scale Discriminative Feature Loss}

Feature-wise (perceptual) loss functions, such as a pre-trained VGG-Net is commonly used as a feature extractor when training image restoration models.
Additionally, adversarial loss is often used as a regularizer to push the solution to the natural image manifold using a discriminator network that is trained to differentiate between distorted and the natural images \cite{ledig2017photo}. However, a fundamental weakness of these methods is that they aim at learning the distribution of natural images using large training datasets, which is less relevant for a loss function. 
In this paper, we introduce our \textbf{Multi-Scale Discriminative Feature (MDF)} loss, which, instead is trained to penalize the task specific distortions that are introduced iteratively to the generator at various stages of training, making the trained discriminator specialized in identifying the distortions for a given restoration method. Unlike VGG and LPIPS networks, which were trained for image classification and the prediction of image quality, respectively, our feature extractor networks are trained for the task that is directly relevant to restoration task in hand. Furthermore, this task is much easier than learning the entire distribution of natural images and thereby can be achieved with as little as a single training image.


The foundations of our loss function are based on the following propositions:

\noindent \textbf{Proposition 1:} \emph{Networks employed as feature extractors for the loss should be trained to be sensitive to the restoration error of the input image. This makes the feature space more suitable for penalizing the distortions during training for that specific task.}

\noindent \textbf{Proposition 2:} \emph{Learning natural-image manifold, which is the task often attributed to discriminators, is a much harder task and is less relevant for the feature-wise loss function. The loss function should be able to detect relevant distortions regardless of image content, i.e. be content invariant.}




To validate both propositions, we design a new feature-wise loss. 
The feature-space comprises the intermediate activations of the set of discriminator networks trained as a single-image GAN \cite{Rott2019} specialized in removing task-specific distortions from a \emph{seed image}. 
We denote the seed image by $\rvy$ to differentiate it from the \emph{training images}, denoted by $\{\rvx_i\}_{i=1}^N$, which are used for learning the restoration task. 
The proposed loss function is trained in a multi-scale manner so that it is sensitive to the relevant distortions at multiple scales. The seed image can have a different size from the training images, can depict a different type of a scene, or can be a synthetic image. Below we revisit the training procedure for our multi-scale discriminators and operation of MDF loss on image restoration tasks.

\subsection{Phase 1: Training the discriminators}

We use the architecture of SinGAN \cite{Rott2019} to train the multi-scale discriminators on a single seed image in a task specific manner. 
A set of generators $\{G^k\}_{k=1}^K$ and a set of discriminators $\{D^k\}_{k=1}^K$ are instantiated for a pre-defined set of $K$ scales. Conventionally, scale $1$ is the coarsest level and scale $K$ is the finest (the original image). In our experiments, we chose $K=8$, resulting in 8 sets of discriminators in the MDF loss. The seed image at scale $k$, $\rvy^k$, is obtained by downsampling (using Lanczos filter) the original image by a factor of $\rho^{(K - k)}$, where $\rho = 2$.


For each scale of training, the generator takes as input the upscaled output of the lower scale after adding the task specific artifacts to it : $\tilde{\rvy}^k = G^k (\tilde{\rvy}^{k-1}\uparrow^\rho +\, \rvz^{k})$. Here $\rvz^{k}$ is the distortion added to the upsampled output of the lower scale $\tilde{\rvy}^{k-1}\uparrow^\rho$. For the first scale of training, the input to the generator is the original image downsampled by a factor of $\rho^{(K - 1)}$ and then distorted by the task-specific error. We have experimented with the generator that was taken directly from SISR network but we did not observe a substantial improvement in performance. Phase 1 of Fig.~\ref{fig:main_diag} provides a graphical illustration of the training scheme.

In contrast to the protocol used for training a single image GAN \cite{Rott2019}, we use application-specific distortion $\rvz^{k}$ while training. For image denoising, $\rvz^{k}$ is Gaussian Noise of a magnitude randomly sampled from a uniform distribution of [0,55] on a pixel scale of (0,255).
For JPEG artifact removal, the upscaled output from the previous scale is compressed with a JPEG quality chosen randomly between 7 and 10, before being fed to the finer scale. 
However, for the task of Single-Image Super Resolution (SISR), no distortion is added ($\rvz^{k}=0$) and the upscaled output from the previous scale is directly fed to the next level.



The corresponding discriminator at scale $k$ takes as input the generated image $\tilde{\rvy}^k$ and produces a map of $[0, 1]$ activations $\tilde{\rvu}^k = D^k (\tilde{\rvy}^k)$ with the same dimensionality as $\tilde{\rvy}^k$. 
Alternatively, the discriminator can be supplied with the downscaled seed image $\rvy^k$, resulting in the activation map $\rvu^k$. The discriminator is trained to distinguish patches of the seed image from patches of the generated image and, therefore, the activations of $\rvu^k$ are pushed towards 1 and those of $\tilde{\rvu}^k$ are pushed towards 0. The number of such activations in the map depends on the number of convolutional layers in the discriminator and their kernel size. In our case, each activation corresponds to an $11{\times}11$ patch in the input. 
Training is done sequentially across scales. The coarsest scale is trained for 3000 iterations, then the weights are frozen and the next scale is trained, and so on. The training loss for the k-th GAN is comprised of an adversarial term and a reconstruction term:
\begin{equation}
    \underset{G^k}{\max} ~ \underset{D^k}{\min} ~ \mathcal{L}_{adv} (G^k,D^k)  + \alpha \mathcal{L}_{rec} (G^k)
\end{equation}
The reconstruction loss $\mathcal{L}_{rec}$ employed is the MSE loss between the generated  $\tilde{\rvy}^k$ and the ground truth image  $\rvy^k$ to ensure faithful generation of the output image. The reconstruction loss weight $\alpha$ is set at 100. Selection of the loss function and the hyper-parameters are based on \cite{Rott2019}.

It must be noted that addition of the above distortions (Gaussian Noise and JPEG compression artifacts) to the \textit{seed image} at various scales makes the discriminator sensitive to such artifacts but agnostic to the image content. The main benefit of our discriminative loss function is that it does not require thousands of images to be trained on, instead a single natural image and knowledge of the distortions are sufficient to provide state-of-the-art results.  

\subsection{Phase 2: Training for image restoration}

In this phase, the trained discriminators are used as the loss function for an image restoration task. For all restoration tasks we use latent embeddings after every ReLU layer of the trained discriminators as features. We denote the embeddings by $\phi^k_l (\rvx)$ meaning the output of the $l$-th layer of the discriminator for the $k$-th scale. The output of the whole discriminator is then $\phi^k_L (\rvx)$, where $L$ is the total number of layers. If $\rvx_i$ is the ground-truth for the i-th training image and $\tilde{\rvx}$ is its reconstruction, then our MDF loss is:
\begin{equation}
\label{eq:loss_function}
    \mathcal{L} = \sum_{k=1}^K ~ \sum^L_{l = 1 } ||\phi^k_l(\rvx) - \phi^k_l(\tilde{\rvx})||^2_2
\end{equation}

A subtle but crucial aspect of our loss is that \textit{the discriminators are not applied to the scales on which they were trained}. If the seed image has dimensions $H_y \times W_y$, the training input (both seed and synthetic) to the discriminator $D^k$ will have dimensions $H_y / \rho^{(K - k)} \times W_y / \rho^{(K - k)}$. However, the input to the discriminator during phase 2 of training will not be scaled and it will be $H_i \times W_i$, the size of the $\rvx_i$. 

\begin{table}[t]

\caption{Comparison of the properties of our proposed loss against other competing losses. 
} 

\label{table:technical-properties}
\centering

\resizebox{.499\textwidth}{!}{
\small
\begin{tabular}{c||c|c|c|c|c|c}

\hline 
 \textbf{Loss function} & \textbf{\makecell{Training \\ overhead}} & \textbf{\makecell{Memory \\ overhead}}  & \textbf{Multi-scale} & \textbf{\makecell{Inference \\ GPU (ms)}} & \textbf{\makecell{Backpropogation \\ time (ms)}} & \textbf{Regularization} \\
\hline \hline
$L_2$ & None & None & No & 1.2 & 1.0 & --\\
$L_1$ & None & None & No & 1.2 &1.0 & -- \\
SSIM \cite{Wang2004} & None & None & No & 12 &1.6 & --\\
MS-SSIM \cite{Wang2003}& None & None & Yes & 24 & 6.5 & --\\
VGG \cite{johnson2016perceptual} & 1.3M images & 58.9MB & No & 27 &21.8  &Required\\
LPIPS \cite{Zhang2018} & 161k images \footnotemark & 9.1MB & No & 31& 17.5 & Required\\
MS-SSIM $+ L_1$ \cite{Zhao2017} & None & None & Yes & 25 &8.2 &--\\
Ours & One image & 4.2MB & No\footnotemark & 11 &4.0 &-- \\

\hline
\end{tabular}
}
\vspace{-0.5cm} 
\end{table}

\footnotetext[1]{This is training on top of a pre-trained network using 1.3M images} 
\footnotetext[2]{Only training of discriminators is performed in a multi-scale fashion.}

\begin{figure*}[t]
    \centering
   {\includegraphics[trim={0cm 1.0cm 0.60cm 0cm}, clip, width=0.99\linewidth]{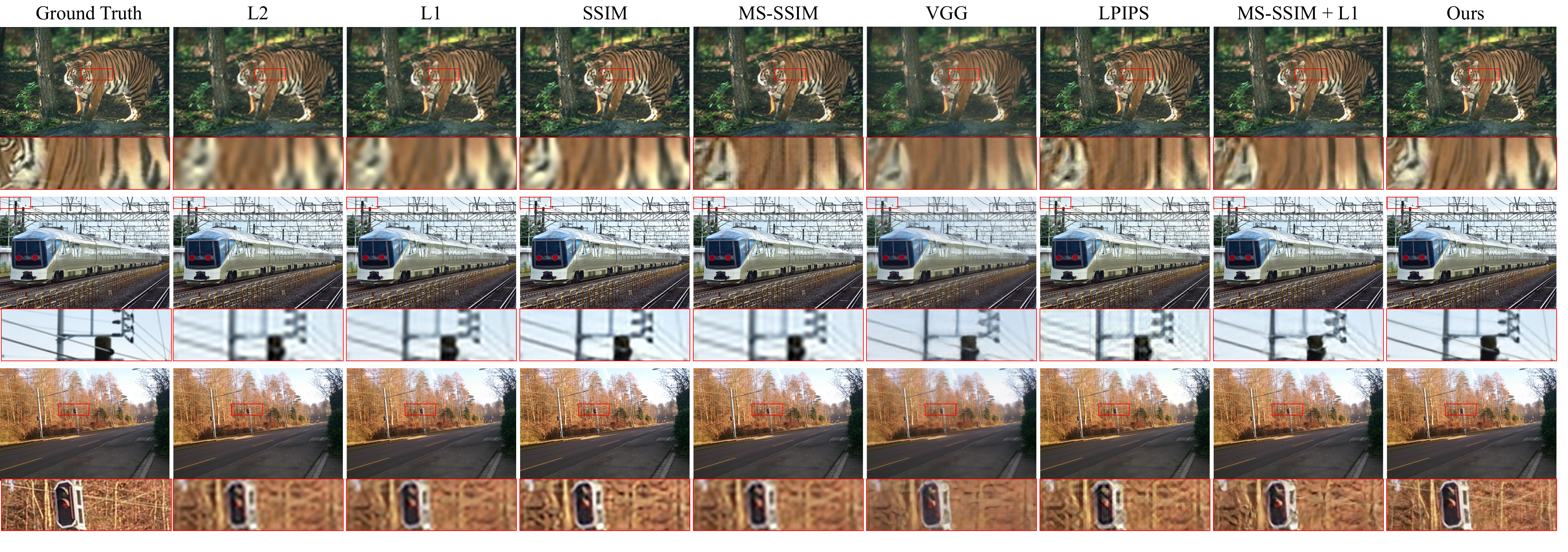} }
    \caption{SISR results for EDSR \cite{lim2017enhanced} trained using different loss functions. Top row shows a sample image from BSD \cite{MartinFTM01}, second row from the DIV2K validation \cite{agustsson2017ntire} and the bottom row from DPED dataset \cite{ignatov2017dslr}. The results for our loss are sharper and have fewer artifacts across all datasets.     
    Best viewed when zoomed. Additional results are provided in the SM.}
    \label{fig:edsr_qualitative}
    \vspace{-0.25cm}    
\end{figure*}

\begin{figure*}[!th]
    \centering
   {\includegraphics[trim={0cm 10.5cm 0.5cm 0cm}, clip, width=0.99\linewidth]{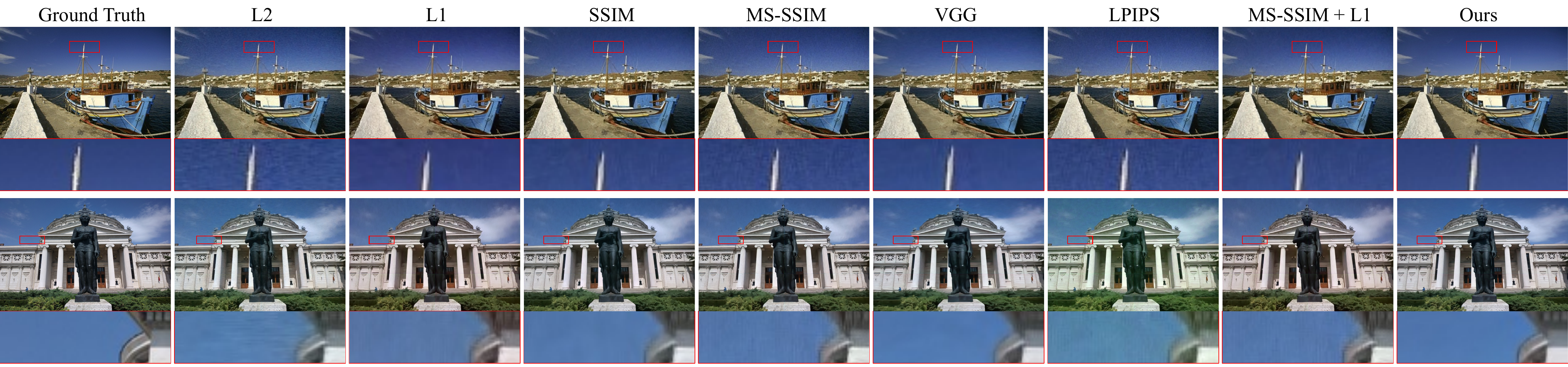} }
    \caption{Results for denoising using DnCNN model \cite{zhang2017beyond} trained using different losses. Top row shows a sample image from BSD \cite{MartinFTM01} and second row from the DIV2K validation \cite{agustsson2017ntire}. Our loss improves noise reduction, especially in the uniform areas of an image. Best viewed when zoomed. Additional results are provided in the SM.
}
    \label{fig:denoising_qualitative}
    \vspace{-0.15cm}    
\end{figure*}

\section{Comparison of loss functions}
\label{sec:evaluatons}
In this section, we evaluate the efficacy of our MDF loss on a variety of image restoration tasks that rely on CNN architectures and also as a regularization term in an adversarial training (Sec.~\ref{sec:gan-training}).
We compare our loss with the most widely used loss functions, listed in Table~\ref{table:technical-properties}, including the perceptual loss \cite{gatys2015texture,johnson2016perceptual}. In all cases, we train the models on the training portion of the DIV2K dataset \cite{agustsson2017ntire} and use for testing DIV2K (the validation set), Berkeley Segmentation Data (BSD 500) \cite{MartinFTM01} and real world mobile phone captured images from the DPED dataset \cite{ignatov2017dslr}. The best model is selected based on the validation loss. 

Note that both VGG and LPIPS losses must be combined with the $L_2$ loss to produce acceptable results. 
For fair comparison, we conducted a hyper-parameter search over the scalar $\lambda$ controlling the weight of the feature-wise loss function. We searched over the values in \{$\lambda: \lambda=10^{k},\,k=-3,..,3$\} for super-resolution and the values of 0.01 and 1 for other applications, due to computational cost. The results of these experiments can be found in the appendix. In our experiments across all restoration applications,
we found the best results are produced when $\lambda = 1$ for VGG and $\lambda = 0.1$ for LPIPS loss. Note that unlike VGG and LPIPS, our MDF loss function does not require addition of $L_2$ regularization while training. It is also worth noting that our MDF loss function is less computationally expensive and has a much lower memory overhead compared to VGG and LPIPS (refer to Table~\ref{table:technical-properties}). 

\begin{figure*}[t]
    \centering
   {\includegraphics[trim={0cm 0cm 0cm 0cm}, clip, width=\linewidth]{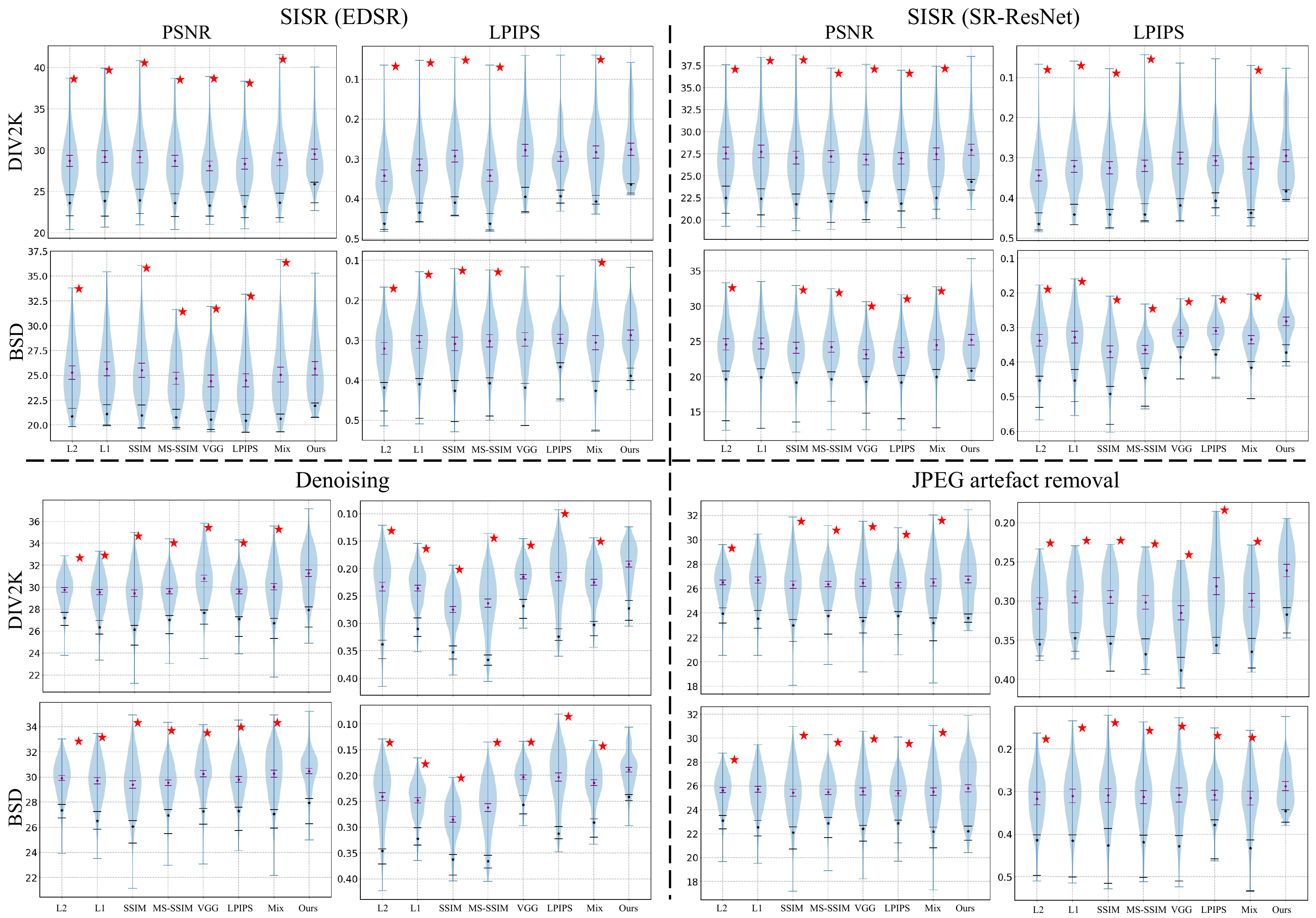} }
   \vspace{-0.6cm}
  \caption{\small{Violin plots illustrating the distribution of the PSNR [dB] $\uparrow$ and LPIPS $\downarrow$ values across different losses across all applications for two datasets. Note that the y-axis is reversed for LPIPS so that the quality improves towards the top of each plot. The error bars show the 95\% confidence intervals for the mean (magenta) and the 5$^{th}$ percentile (black). The latter CIs were computed by bootstrapping. The red asterisks indicate that one-tailed t-test on the means gives statistically significant difference at $\alpha=0.05$. It is worth noting that our loss produced fewer images with low quality values.}}
  \vspace{-0.4cm}
  \label{fig:psnr_lpips_violin_plots}  
\end{figure*}

\paragraph{Single-image super resolution}
\label{subsec:SISR}
Here, we evaluate our proposed loss for the task of SISR, which aims at estimating a High-Resolution (HR) image from a given Low-Resolution (LR) image. For SISR, we use two state-of-the-art architectures, namely Enhanced Deep Super-Resolution (EDSR) \cite{lim2017enhanced} and SR-ResNet \cite{ledig2017photo}. The LR image is generated by downsampling the original HR image by a factor of 4 using bicubic filter. For training, we randomly extract $96\times96$ patches from the dataset and perform data augmentation with 90$^\circ$, 180$^\circ$ and 270$^\circ$ rotations, and horizontal and vertical flips. Each model is trained for 500 epochs with an initial learning rate of 0.001 with gradual rate scheduling.

\paragraph{Image denoising}
\label{subsec:image_denoising}

We train the DnCNN architecture proposed by Zhang \textit{et al.} \cite{zhang2017beyond}. The training set is generated by adding Gaussian noise with the standard deviation randomly selected from the uniform distribution of [0,55]. We use SGD with a weight decay of 0.0001 with Nestrov momentum optimizer for training. Each model is trained for 50 epochs with an exponential learning rate scheduling from 0.1 to $10^{-4}$ with the momentum parameter set to 0.9. 



\paragraph{JPEG artifact removal}
For this application, we use the same DnCNN \cite{zhang2017beyond} as for the denoising.
During training we feed in images compressed with the JPEG codec with the quality factor of 10 as in \cite{dong2015compression,galteri2017deep}. 
We perform data augmentation with 90$^\circ$ image rotation, vertical and horizontal flips. The model is trained with Adam optimizer and the learning rate set to $1e-4$. The test images are compressed with a quality factor of 10 and a more challenging factor of 7. 

\subsection{Qualitative results}

In Figs.~\ref{fig:edsr_qualitative} and~\ref{fig:denoising_qualitative} we provide qualitative results for SISR and image denoising respectively. The examples for other applications can be found in the appendix. Furthermore, we include an extensive set of results at the original resolution in a separate HTML report. The visual results consistently indicate that our task-specific loss can produce sharper, less noisy images with fewer artifacts. The differences are the most noticeable in the flat areas of the images. 


\subsection{Quantitative results}

The quantitative results for all four applications are shown as distributions in
Fig.~\ref{fig:psnr_lpips_violin_plots} for two test datasets: DIV2K and BSD. We report the quantitative results in tabular form and provide additional results for DPED dataset in the appendix. The differences in means (magenta dots in Fig.~\ref{fig:psnr_lpips_violin_plots}) are small but statistically significant for most comparisons (one-tailed t-test with $H_1$ show that the quality score is higher for our method, red $*$ symbols are shown if the difference is significant at $\alpha=0.05$). The means, however, are not the best indicator of performance of different losses. This is because the differences in loss functions are mostly visible in smooth or flat parts of the images, which occupy only small percentage of all pixels but have a substantial impact on the perceived image quality (as demonstrated in Sec.~\ref{subsec:user-study}). The advantage of our loss is better visible for the worst-case results, shown in Fig.~\ref{fig:psnr_lpips_violin_plots} as the lower 5$^{th}$ percentile of values (black asterisks). In majority of the comparison, MDF loss produces fewer images with low quality values, especially in terms of LPIPS.

\subsection{Subjective quality assessment}
\label{subsec:user-study}


\begin{figure}[t]
    \centering
   {\includegraphics[trim={4cm 1.2cm 4.33cm 0.6cm}, clip, width=\linewidth]{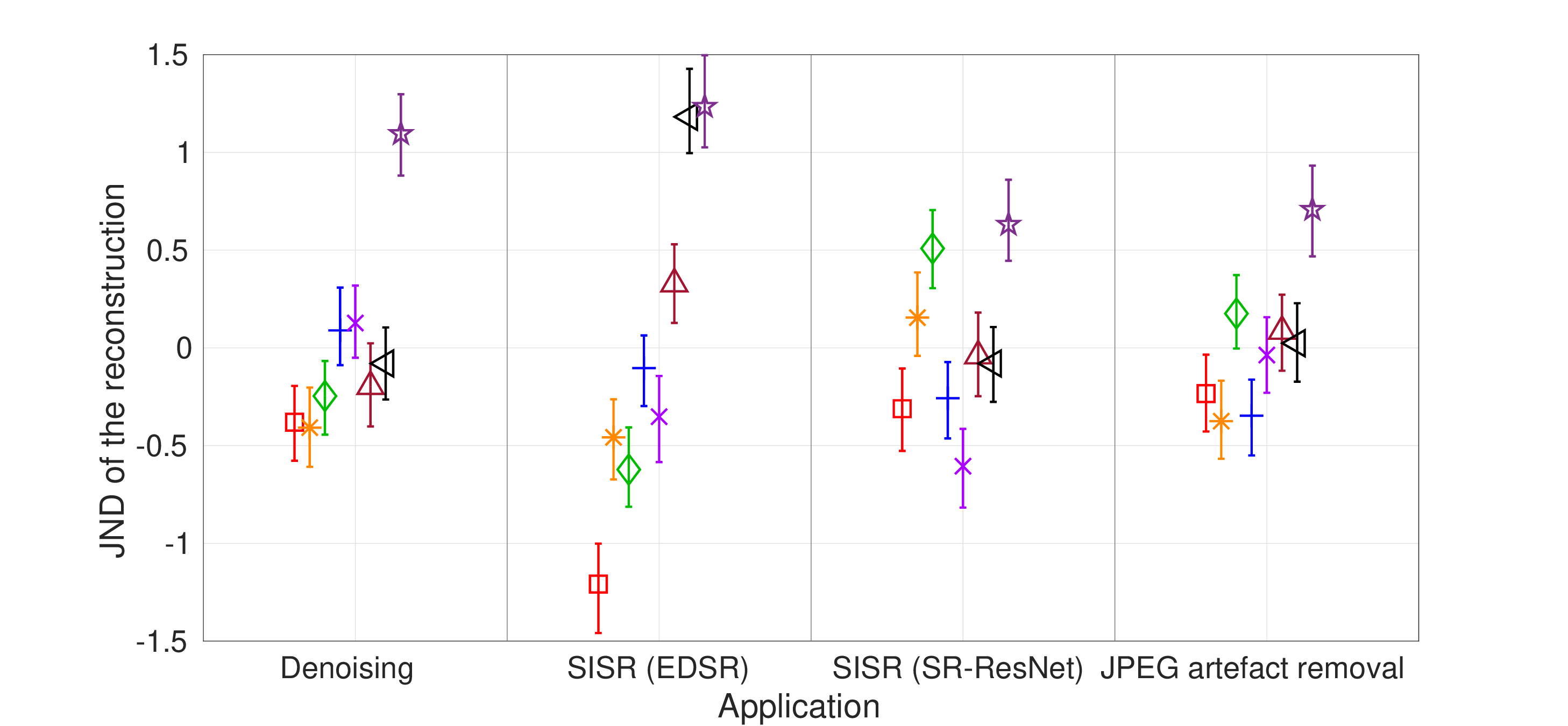} }
   \vspace{-0.6cm}
  \caption{\small{Subjective experiment in JND units (the higher, the better). 
  Error bars denote 95\% confidence intervals. The legend is same as Fig.~\ref{fig:perc_dist_tradeoff}. 
  }}
  \label{fig:subjective_experiment}  
  \vspace{-0.15cm}
\end{figure}


Objective metrics such as PSNR or LPIPS, can be unreliable in predicting the perceptual quality of images. They also do not capture the practical significance of the perceptual difference; we do not know whether the improvement of 0.5\,dB is going to be appreciated by an average observer. For that reason, we ran perceptual experiments on the Amazon Mechanical Turk crowd-sourcing platform. 



For best sensitivity of the test, we used full-design pairwise-comparison protocol \cite{perez2017}. In each trial, participants were presented with 3 side-by-side images: one reference and two generated by the image reconstruction methods, each with different loss function from the BSD dataset. Participants were asked to select the image that appeared closer to the reference. For fair comparison, we randomly selected 50 images from testset. Thus, every loss function was compared to every other loss 50 times. Overall, we collected 1400 comparisons for each restoration method. 

In each Human Intelligence Task (HIT) we included nine pairwise comparison trials and one (for denoising and JPEG artifact removal) or two (EDSR and SR-ResNet) additional pairwise comparisons with an obvious outcome to screen the results against the participants who misunderstood the task. If a participant made a mistake in those comparisons, we excluded that HIT. Overall we discarded 4.2\% and 14.2\% comparisons for SISR with EDSR and SR-ResNet respectively, 7.1\% comparisons for denoising and 9.2\% comparisons for JPEG artifact removal.


  

\begin{figure}[t]
    \centering
   {\includegraphics[trim={0cm 0cm 0cm 0cm}, clip, width=\linewidth]{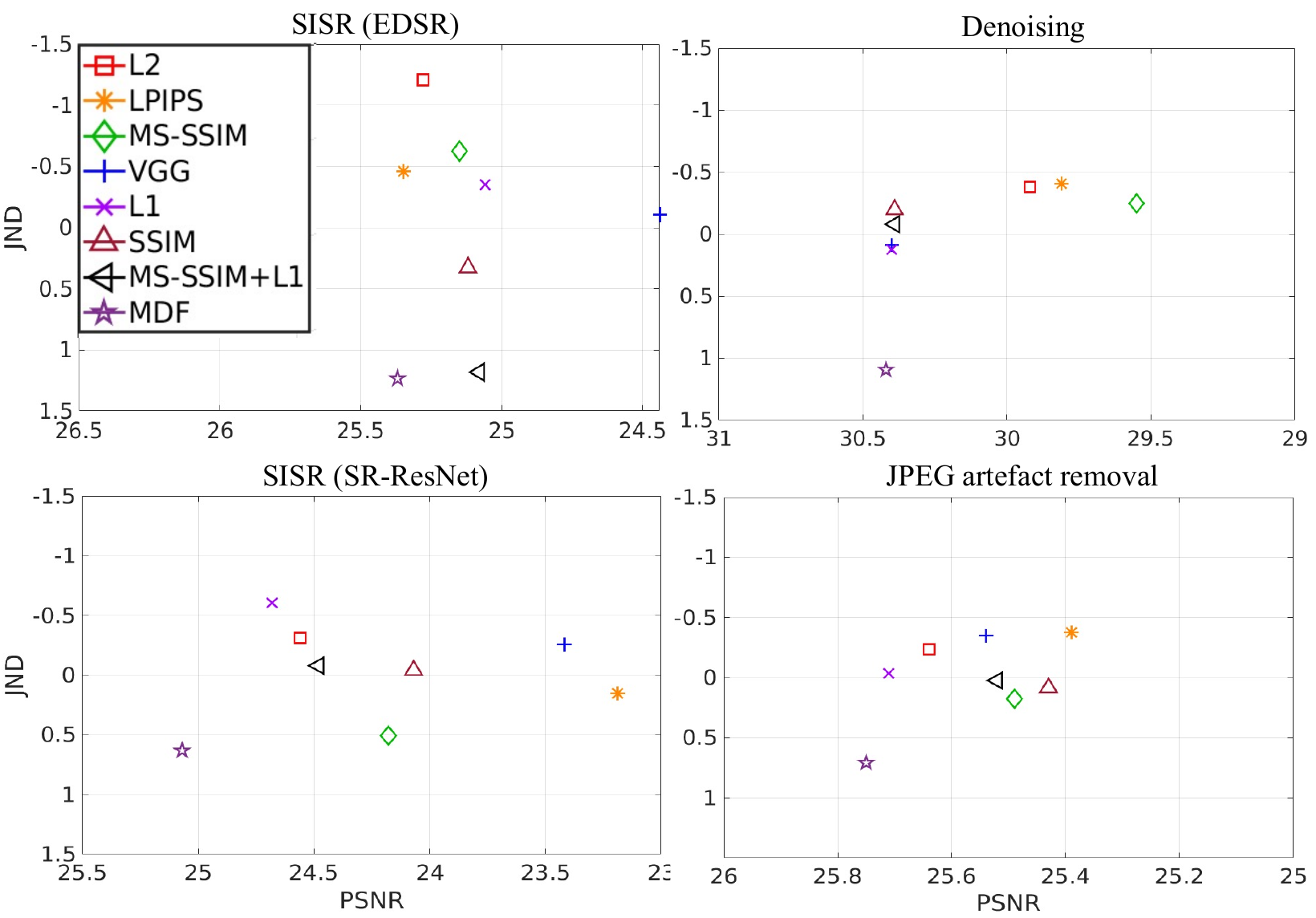} }
   \vspace{-0.6cm}
  \caption{\small{Perception-distortion trade-off for the tested losses. The axes have been reversed so that the lowest distortion is shown on left and the highest perceptual quality at the bottom as in \cite{Blau2018CVPR}.} 
  }
  
  \label{fig:perc_dist_tradeoff}  
   \vspace{-0.5cm}
\end{figure}

For each application we aggregated collected comparisons and performed Just Noticeable Difference (JND) (Thurstonian) scaling on the results using the method from \cite{perez2017}. 
The results express the quality difference in JND units. One JND unit means that 75\% of the population will select one method over another (from a pair). The results of the scaling, plotted in Fig.~\ref{fig:subjective_experiment}, show consistent improvement of our method over other losses. MS-SSIM + L$_1$ performed the second best for SISR on the EDSR, with MDF having an advantage of 0.05~JND. For other applications, MDF shows a substantial improvement over all competing losses. 



To gain further insights, in Fig.~\ref{fig:perc_dist_tradeoff} we visualize the results as the perception-distortion trade-off \cite{Blau2018CVPR}, which shows the distortion (PSNR) on the x-axis and the JND quality values on the y-axis (reversed scale). The results across all applications clearly show that the proposed MDF loss results in both the lowest distortion and the highest perceived quality. The results for EDSR show drastic difference in the performance as measured by PSNR and subjective experiment. MDF and $L_2$ -- the best and the worst performing losses, differ only by 0.09 PSNR, but have 2.4 JND difference in the perceptual quality, corresponding to $94.7\%$ of the population selecting the results produced by MDF.





\subsection{Comparison with adversarial loss}
\label{sec:gan-training}
Our MDF loss can be also used as a reconstruction term when training a GAN architecture for image restoration. 
For this experiment, we chose the task of SISR and used state-of-the-art GAN based method --- ESRGAN \cite{Wang_2018_ECCV_Workshops}. The model trained with the MDF loss function alongside the adversarial loss  achieves a PSNR of 25.37\,dB as compared to the weighted combination of VGG and MSE loss function's 25.06\,dB. Both the models are trained using the DIV2K dataset and inference is run on the BSD dataset. Since, models trained with adversarial loss are known to produce lower PSNR values, we further conducted a subjective study to predict the perceptual quality of the images. We ran a pairwise comparison study on 50 randomly selected images from the testing dataset with each pairwise comparison performed four times. MDF has an advantage over VGG loss function and was selected in 58\% of the comparisons.

\begin{figure*}[t]
    \centering
   {\includegraphics[trim={0cm 0cm 0cm 0cm}, clip, width=0.95\linewidth]{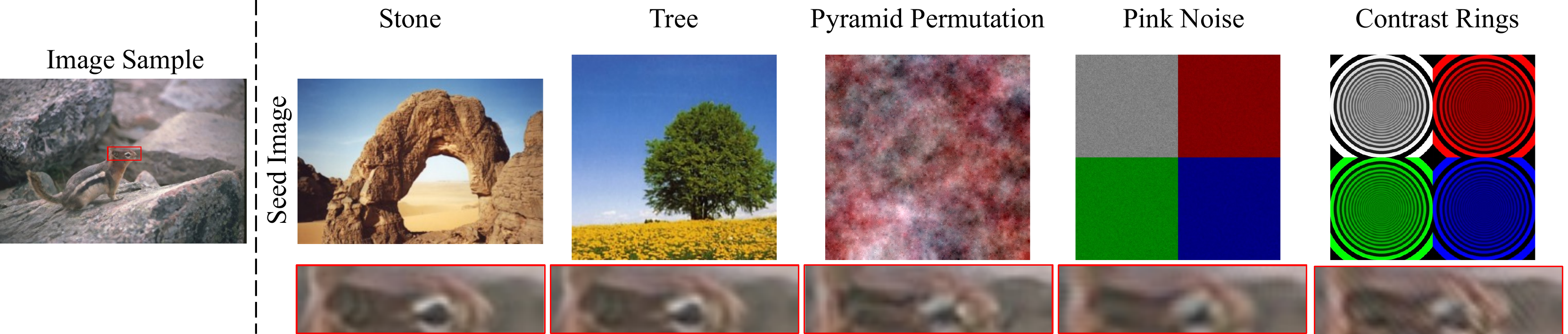} }
   \vspace{-0.25cm}
    \caption{Ablation study on changing the image used for training our MDF loss function. It can be seen that natural images provide visually better results as compared to synthetic images. Best viewed when zoomed.}
    \label{fig:ablation-diff-images}
    \vspace{-0.25cm}    
\end{figure*}

\section{Ablation analysis}
\label{subsec:ablation}


The ablation studies test the importance of task-specific MDF, the choice of seed image, the number of images used to train the discriminator and the number of discriminator scales. The latter two studies are described in the appendix. 

\begin{figure}[t]
    \centering
   {\includegraphics[width=\linewidth]{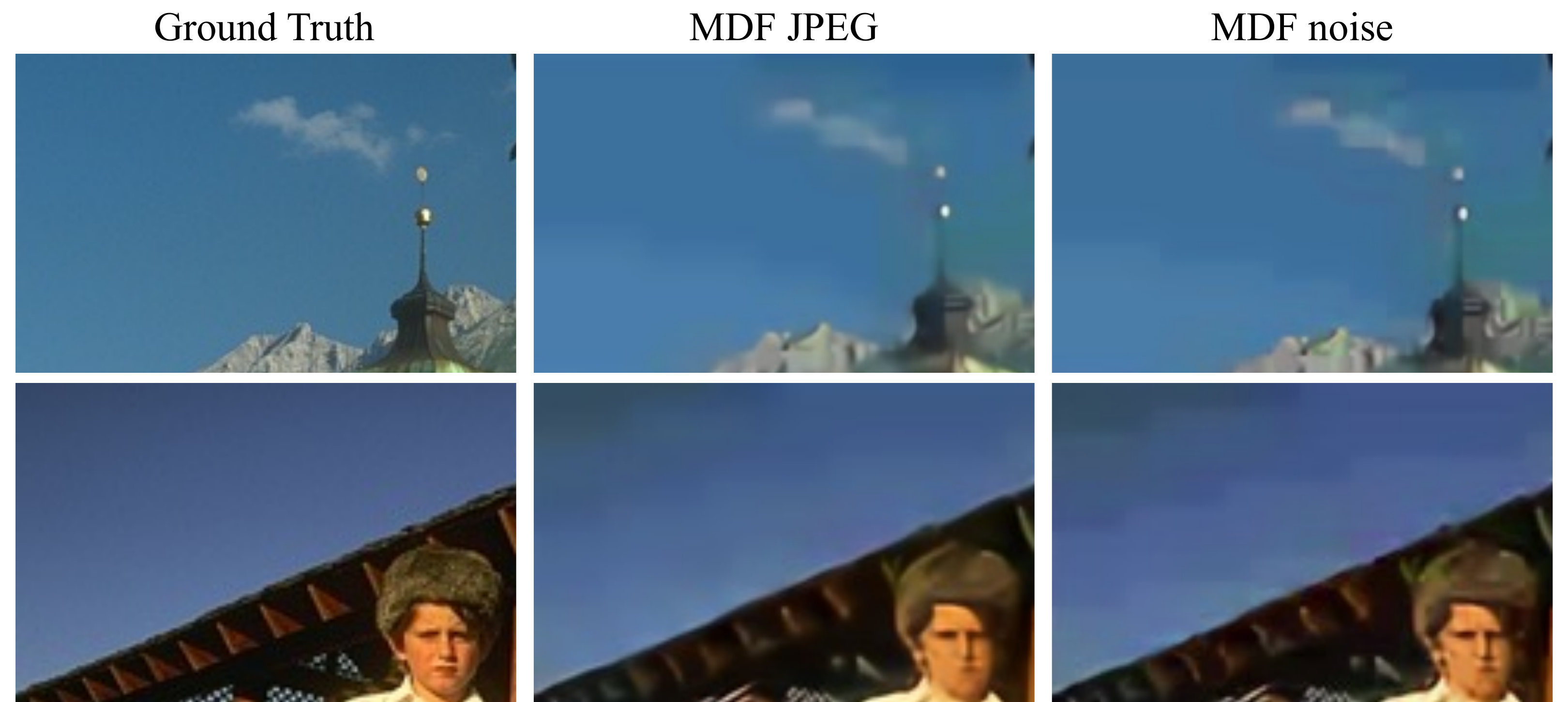} }
   \vspace{-0.45cm}
  \caption{Example results for JPEG artifact removal when trained on task-specific MDF~JPEG and MDF~noise. Task-specific MDF results in improved visual quality. 
  }
  \label{fig:ablation_jpeg}  
  \vspace{-0.4cm}
\end{figure}

\paragraph{Task-specific distortions}
Here we test whether the loss trained on one task can serve as a feature extractor for another. We train DnCNN for JPEG artifact removal using MDF with introduced either noise or JPEG distortions (different vectors $\rvz^{k}$). The task-specific discriminator gained moderate performance increase in terms of PSNR (25.75\,dB for MDF~JPEG and 25.61\,dB for MFD~noise) but resulted in images of much better visual quality, as shown in Fig.~\ref{fig:ablation_jpeg}. 

\paragraph{Seed image}
We study the effect of using different natural and synthetic images for training our MDF loss function. Fig.~\ref{fig:ablation-diff-images} shows 5 seed images including 2 natural and 3 synthetic ones that were used to train the discriminators. \emph{Pyramid Permutation} image has been created by a random permutation of pixel order on each level of the Laplacian pyramid. Such permutation distorts image second-order statistics, but preserves the composition of the spatial spectrum. \emph{Pink Noise} image contains $\nicefrac{1}{f^2}$ noise that is typical for natural images. \emph{Contrast Rings} image contains concentric rings whose contrast is reduced towards the centre to cover the range of edges of all orientations and contrast magnitudes. The results of SISR (EDSR), shown in the bottom part of Fig.~\ref{fig:ablation-diff-images}, indicate that the visual quality of the super-resolved images is best for natural images and is degraded as the statistics of the training image is distorted. However, from the results for all the applications, the visual quality of the restored images is more dependent on the nature of distortions added ($\rvz^{k}$) than the choice of the seed image.


\section{Image quality metrics and loss functions}

\begin{SCfigure}
\includegraphics[trim={1.40cm 0cm 1.75cm 0.85cm}, clip, width=0.29\textwidth]{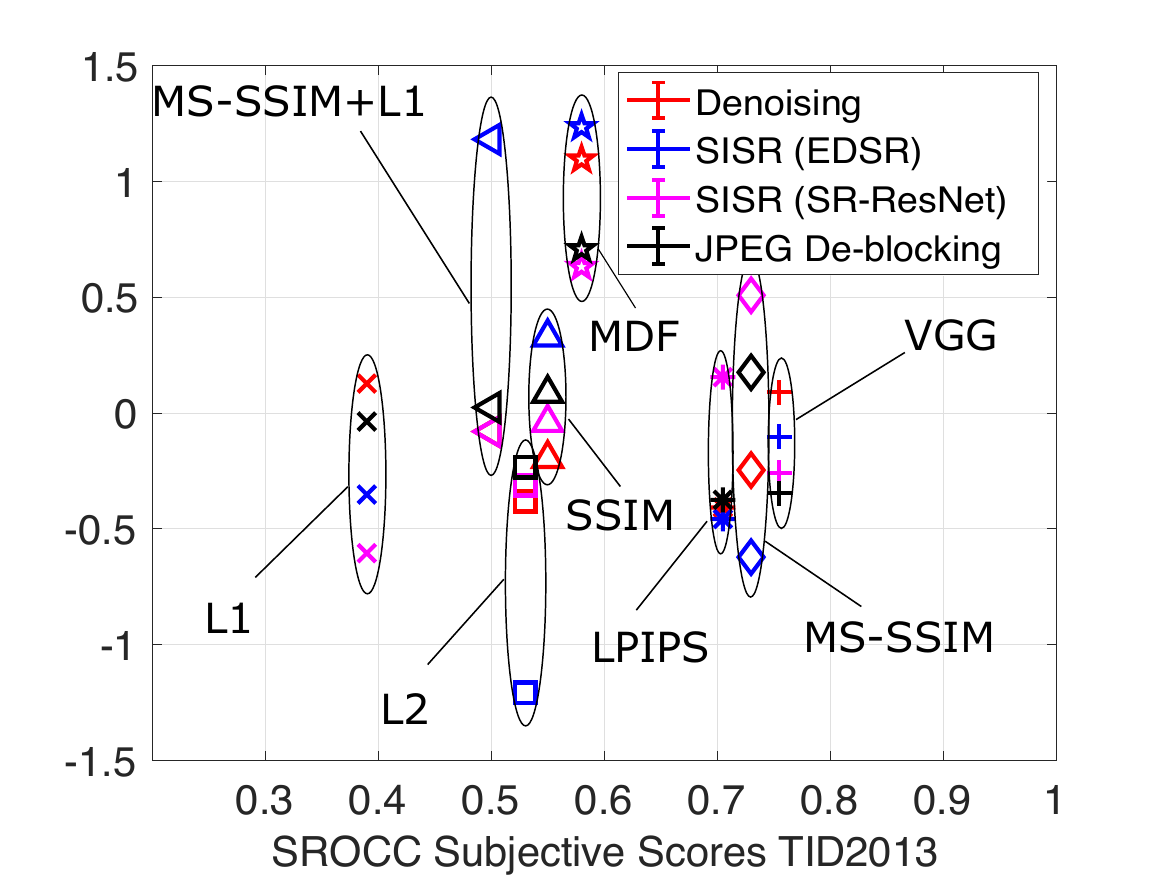}
\caption{\small{Performance of loss functions on the task of image quality prediction versus performance as objective functions. Results do not show strong correlation. Markers are consistent with Fig.~\ref{fig:subjective_experiment}}}
\label{fig:srocc_all}
\vspace{-0.3cm}
\end{SCfigure}

Provided with the results of our subjective quality experiment from the previous section, we further test whether a good loss function is also a good image quality metric.  Here, we used each loss function as a quality predictor for the improved version of the TID2013 dataset \cite{Mikhailiuk2018}. The dataset is one of the most accurate (due to large number of comparisons), is scaled in JND units, and contains sufficiently large number of conditions (over 4000 images). In Fig.~\ref{fig:srocc_all} we plot the Spearman Rank Order Correlation Coefficient (SROCC) with the subjective scores from the improved TID2013 against the JND values from our subjective experiment. High SROCC value indicate that the loss is a good predictor of image quality. The scatter plot shows little correlation; the best quality predictors are not necessarily the best loss functions. This is an important finding because it puts in question whether loss functions should be optimized for prediction of image quality. Additional experiments to investigate the performance of various loss functions as quality predictors are provided in the appendix.

\section{Conclusions}

In this paper, we have shown several observations that go against the common assumptions of what make a good loss. We demonstrated that a small multi-scale discriminator network, trained to detect application-specific distortions, can serve as a better feature-wise loss than large networks, such as VGG, which have been trained on large datasets. This shows that learning a natural image manifold, semantic, or style features may not be essential for an effective loss function. Instead, the loss needs to penalize errors specific for restoration task in hand. Our subjective assessments reveal that the restored images generated using models trained with a task specific loss function are consistently chosen by human observers to be closer to the reference images.

\newpage

\section*{\Large Appendix}
\setcounter{section}{0}

This appendix includes additional details that could not be included in the main paper due to the lack of space. This comprises: \textit{a)} manifold assumption validation and visual comparison with SR-GAN discriminator as compared to our multi-scale discriminators; \textit{b)} quantitative results in terms of average PSNR and LPIPS for all application across 3 benchmark datasets \textit{c)} qualitative results for the JPEG artefact removal application; \textit{d)} ablation study on the number of seed images and the number of discriminators of the Multi-Scale Discriminative Feature (MDF) loss; \textit{e)} hyper-parameter tuning for the VGG and LPIPS feature-wise loss functions; and \textit{f)} performance of loss functions as quality predictors.

\section{Image manifold assumption}
\label{sec:manifold}

The main objective of GANs \cite{Goodfellow2014} in image restoration is to learn a discriminator model that differentiates between image manifolds \cite{li2016combining,yeh2016semantic,mustafa2020transformation,mathieu2015deep,denton2015deep}. This is based on the hypothesis that input samples (e.g. noisy images) and their corresponding ground truth samples lie on two different manifolds. The generator model thereby learns a mapping function from one manifold to another, resulting in photo-realistic images closer to the natural image manifold \cite{ledig2017photo,denton2015deep}. 

However, in this paper, we propose that learning the natural image manifold, which is often the task attributed to the discriminator, is less important than being able to detect errors introduced by the generator. Moreover, learning the natural image manifold requires the GAN to be trained with thousands of natural and fake images, making the training process computationally intensive. Here, we show that our task-specific discriminators, trained on a single image, can be used as feature extractors for the loss function because they learn the generator errors rather than the natural image manifold. 

\begin{figure}[t]
    \centering
  {\includegraphics[trim={0cm 0cm 0cm 0cm}, clip, width=0.98\linewidth]{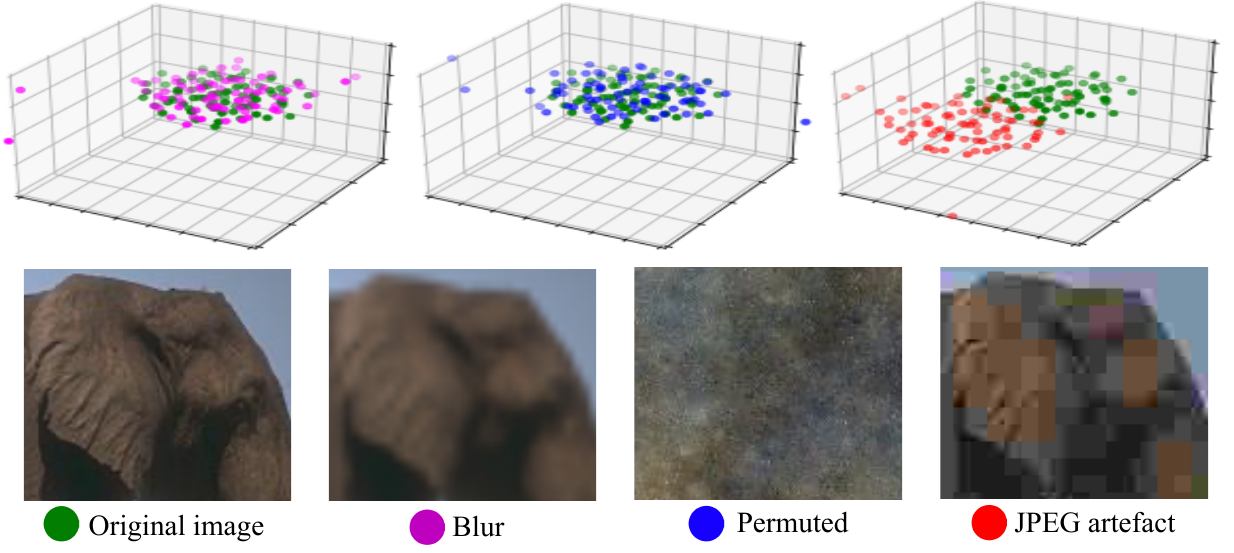} }
    \caption{Manifold assumption validation: The figure shows the 3D t-SNE plots of the latent feature vectors extracted from diverse sets of images using multi-scale discriminators trained for the JPEG artefact removal task. Our JPEG-tuned discriminator cannot differentiate between the original and permuted images (middle plot), yet is a very effective feature-extractor for a loss function for JPEG task. }
    \label{fig:manifold}
\end{figure}

To validate this claim, a multi-scale discriminator trained on \emph{a single image} for the task of JPEG artefact removal is employed as feature extractor. We randomly sample 100 natural images from the ILSVRC validation dataset \cite{Russakovsky2015}. From these images we generate \textit{a)} JPEG compressed images using a compression quality between 7 and 10, \textit{b)} blurry image samples by downsampling and upsampling the images by a factor of 4 using bi-linear filter and \textit{c)} scrambled images by randomly permuting the pixels on each level of the Laplacian pyramid. Such permutations distort the second-order statistic, but preserve the composition of the spatial spectrum. The JPEG trained discriminator is used to extract the latent feature space of each set of images. The feature space for each image is the average across the channels and the resulting feature vector is reduced to a dimensionality of 3 using t-SNE for visualization. Fig.~\ref{fig:manifold} shows the plot of the features from each set of images. The visualization shows that the discriminator does not learn the natural image manifold and cannot discriminate between natural and randomly permuted images. It also cannot discriminate between blurred and original images, but performs well in detecting JPEG artifacts regardless of image content. 

\subsection{Image manifold comparison}
In this section, we repeat the experiment conducted above, instead this time for a fully trained SR-GAN \cite{ledig2017photo} discriminator. This further bolsters our claim that the task-specific discriminators of our MDF loss function learn to detect the generator distortions instead of the entire natural image manifold. This thereby allows our MDF loss function, trained on a single image, to be used to effective feature extractors between the generated and the reference image.

\begin{figure}[t]
    \centering
  {\includegraphics[trim={0cm 0cm 0cm 0cm}, clip, width=0.98\linewidth]{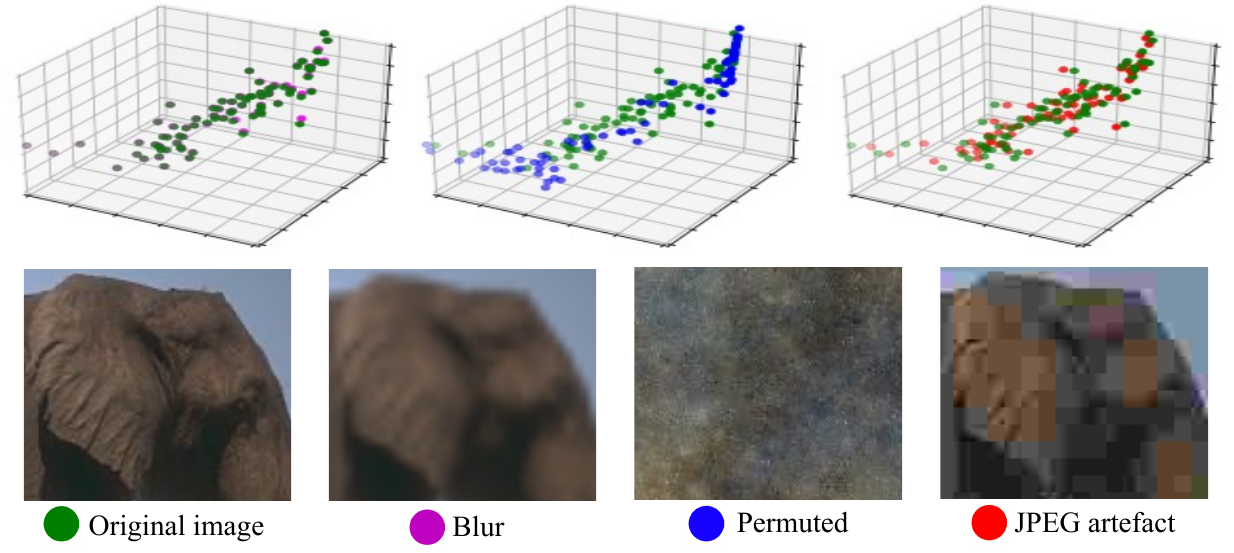} }
    \caption{Manifold assumption validation: The figure shows the 3D t-SNE plots of the latent feature vectors extracted from diverse sets of images using an SR-GAN discriminator trained on DIV2K dataset \cite{agustsson2017ntire}. The SR-GAN discriminator cannot differentiate between the original and jpeg images (right plot), thereby cannot be used as an effective feature extractor to detect and remove distortions. }

    \label{fig:manifold_srgan}
\end{figure}

We chose the same sample of 100 natural images from the ILSVRC validation dataset \cite{Russakovsky2015}. From these images we generated a) JPEG compressed images using a compression quality between 7 and 10, b) blurry image samples by downsampling and upsampling the images by a factor of 4 using bi-linear filter and c) scrambled images by randomly permuting the pixels on each level of the Laplacian pyramid. Such permutations distort the second-order statistic, but preserve the composition of the spatial spectrum. A trained SR-GAN discriminator is used to extract the latent feature space of each set of images. The feature space for each image is chosen after the Global Average Pooling (GAP) layer of the network. We used t-SNE to reduce the dimensionality of the feature vector to 3 for visualization. Fig.~\ref{fig:manifold_srgan} shows the plot of the features from each set of images. The visualization shows that the discriminator of SR-GAN learns the natural image manifold (unlike our multi-scale discriminator) and can discriminate between natural and randomly permuted images. However, it cannot discriminate between the JPEG compressed and original images, making it an inferior feature extractor to detect and remove distortions.

\section{Quantitative results}

\begin{figure*}[!th]
    \centering
   {\includegraphics[trim={0cm 0cm 0cm 0cm}, clip, width=0.995\linewidth]{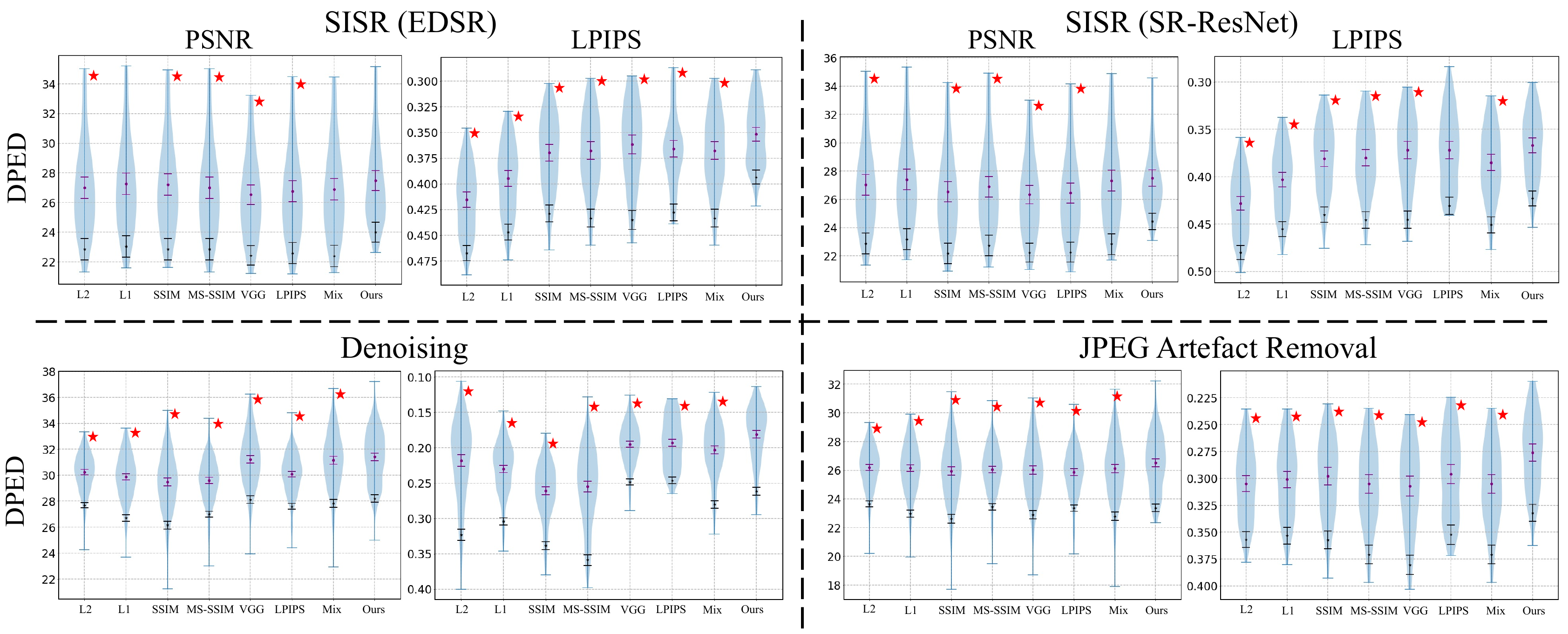} }
    \caption{Additional violin plots illustrating the distribution of the PSNR [dB] $\uparrow$ and LPIPS $\downarrow$ values for DPED dataset \cite{ignatov2017dslr} for all applications. Note that the y-axis is reversed for LPIPS so that the quality improves towards the top of each plot. The error bars show the 95\% confidence intervals for the mean (magenta) and the 5$^{th}$ percentile (black). The latter CIs were computed by bootstrapping. The red asterisks indicate that one-tailed t-test on the means gives statistically significant difference at $\alpha=0.05$. It is worth noting that our loss produced fewer images with low quality values.
}
    \label{fig:Violin_Plots_Supplementary}
\end{figure*}

The quantitative results for all four applications are shown as distributions in Fig.~\ref{fig:Violin_Plots_Supplementary} for real world mobile phone captured images from DPED dataset \cite{ignatov2017dslr}. The differences in means (magenta dots in Fig.~\ref{fig:Violin_Plots_Supplementary}) are small but statistically significant for most comparisons (one-tailed t-test with $H_1$ show that the quality score is higher for our method, red $*$ symbols are shown if the difference is significant at $\alpha=0.05$). The means, however, are not the best indicator of performance of different losses. This is because the differences in loss functions are mostly visible in smooth or flat parts of the images, which occupy only small percentage of all pixels but have a substantial impact on the perceived image quality (as demonstrated in Sec.~4.3 of the main paper). The advantage of our loss is better visible for the worst-case results, shown in Fig.~\ref{fig:Violin_Plots_Supplementary} as the lower 5$^{th}$ percentile of values (black asterisks). In majority of the comparison, MDF loss produces fewer images with low quality values, especially in terms of LPIPS. 
We also report the quantitative results in terms of average PSNR and LPIPS in Table.~\ref{table-various-loss-comparison}.

\section{JPEG artefact removal results}
\begin{figure*}[!th]
    \centering
   {\includegraphics[trim={0cm 1.95cm 0.5cm 0cm}, clip, width=0.995\linewidth]{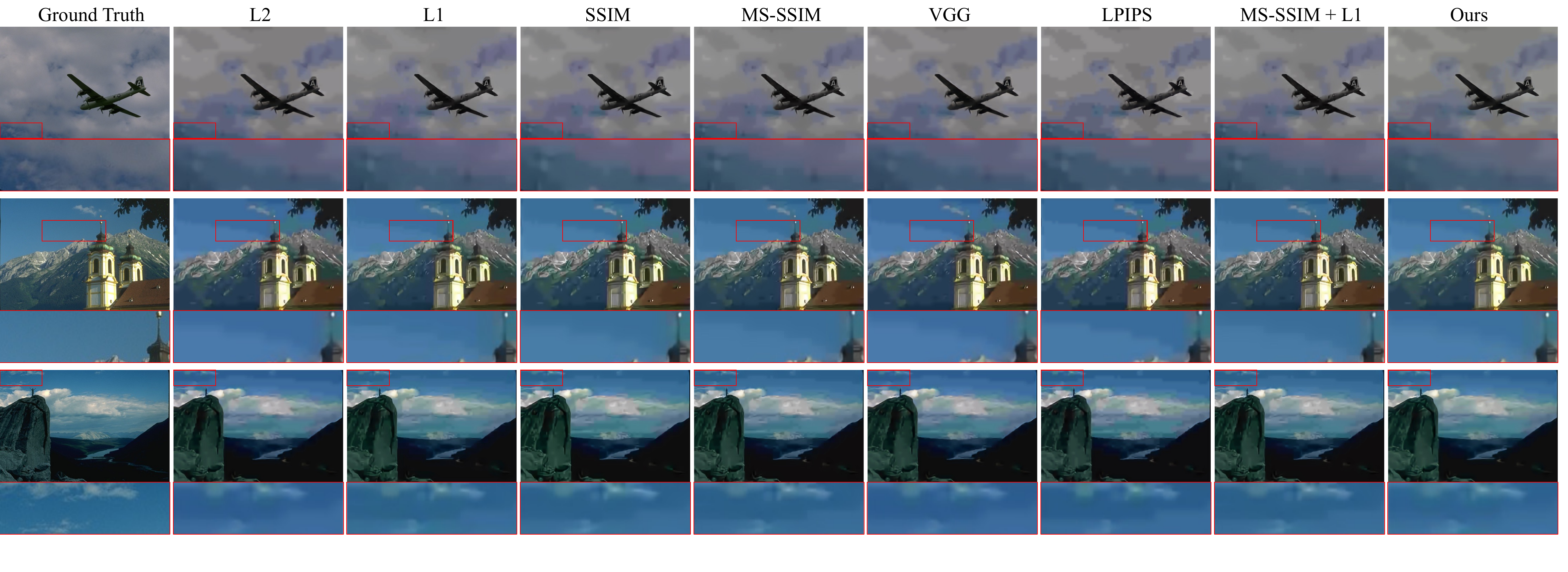} }
    \caption{Results for JPEG artefact removal (compression quality = 7) using DnCNN model \cite{zhang2017beyond} trained using different losses. Our loss improves artefact reduction, especially in the uniform areas of an image. Qualitative results in terms of PSNR and LPIPS are reported in Table~\ref{table-various-loss-comparison}. Best viewed when zoomed. 
}
    \label{fig:jpeg_7}
\end{figure*}

In this section, we provide qualitative results showing comparison between three sample reconstructed images from the BSD Test Set using our (MDF) loss with various other loss functions for the task of JPEG artefact removal application. The test images are compressed with a quality factor of 10 and a more challenging factor of 7. Fig.~\ref{fig:jpeg_7} shows the results for the compression quality factor 7. The performance of the various loss functions seems to be comparable for the quality factor of 10, however, our model substantially provides artefact removal, especially in the uniform areas of the image for a much challenging codec quality of 7. The same was also observed in the subjective experiment conducted (see Sec.~4.3 of the main paper).

\section{Ablation study}
\subsection{Scales of Discriminators}
Since our MDF loss function comprises a series of discriminators trained on a single image at various scales, we need to select the optimal number of scales (the hyper-parameter $K$ in Equation~2 of the main paper) to achieve the best performance. We perform an ablation study on training the EDSR model  \cite{lim2017enhanced} using only the coarsest scale discriminator and subsequently adding finer scales. We observe a significant increase in quality of the images generated with the increase in the number of discriminators. As shown in Table~\ref{table-ablation-scales}, our loss performs the best when all 8 scales are employed. 

\begin{table}[t]

\caption{Ablation study on training the SISR model (EDSR) using different scales of our loss. The scale number represents the number of scales included in the MDF loss. The inference results are reported for the BSD dataset.} 

\label{table-ablation-scales}
\centering

\scriptsize
\begin{tabular}{|c|c|c|c|c|c|c|}

\hline \rowcolor{black!10}
 Scales & 1 & 2 & 3  & 5  & 7 & 8\\
\hline \hline
PSNR $\uparrow$ & 22.55 & 23.89 & 24.43 & 24.89 & 25.27  & \textbf{25.70} \\

LPIPS $\downarrow$ &0.392 & 0.357& 0.354& 0.311 & 0.305 & \textbf{0.286}\\

\hline
\end{tabular}
\end{table}

\begin{table*}[!h]
\small
\caption{Comparison of our proposed Multi-Scale Discriminative Feature (MDF) loss function with other losses on 3 public benchmark datasets for four tested applications. Results show PSNR [dB] $\uparrow$ / LPIPS $\downarrow$. The numbers in red indicate the best performance and the ones in blue the second best. 
} 
\label{table-various-loss-comparison}
\centering

\resizebox{.98\textwidth}{!}{

\begin{tabular}{c||c|c|c|c|c|c|c|c}

\hline \rowcolor{black!10}

 \multirow{1}{*}{\textbf{Dataset}} & \multirow{1}{*}{\textbf{L$_2$}} & \textbf{L$_1$}  & \textbf{SSIM}   & \textbf{MS-SSIM} & \textbf{VGG} & \textbf{LPIPS} & \textbf{MS-SSIM + L$_1$} & \textbf{Ours}\\
\hline \hline

 \multicolumn{9}{c}{Single Image Super-Resolution (EDSR \cite{lim2017enhanced})}\\
\hline
DIV2K  & 28.70 / 0.342 & \blue{29.22} / 0.315 & 29.21 / 0.293 & 28.70 / 0.342 & 28.10 / \blue{0.278} & 28.34 / 0.283 & 28.87 / 0.283 & \red{\textbf{29.51}} / \red{\textbf{0.276}}\\
DPED  & 26.99 / 0.415 & \blue{27.26} / 0.394 & 27.22 / 0.369 & 27.00 / 0.367 & 26.54 / \blue{0.361} & 26.76 / 0.366 & 26.88 / 0.368 & \red{\textbf{27.48}} / \red{\textbf{0.351}}\\
BSD    & 25.28 / 0.320 & \blue{25.66} / 0.304 & 25.52 / 0.309 & 24.70 / 0.301 & 24.44 / 0.298 & 24.49 / \blue{0.296} & 25.08 / 0.306 & \red{\textbf{25.70}} / \red{\textbf{0.286}} \\

\hline \hline
\multicolumn{9}{c}{Single Image Super-Resolution (SR-ResNet \cite{ledig2017photo})}\\
\hline
DIV2K & 27.57 / 0.343 & \blue{27.76} / 0.321 & 27.05 / 0.325 & 27.20 / 0.320 & 26.83 / \blue{0.301} & 27.00 / 0.307 & 27.49 / 0.313 & \red{\textbf{27.95}} / \red{\textbf{0.295}} \\
DPED  & 27.03 / 0.428 & \blue{27.41} / 0.403 & 26.54 / 0.381 & 26.89 / 0.380 & 26.34 / \blue{0.372} & 26.45 / \blue{0.372} & 27.32 / 0.385 & \red{\textbf{27.50}} / \red{\textbf{0.367}}\\
BSD & 24.56 / 0.337 & \blue{24.68} / 0.328 & 24.07 / 0.370 & 24.18 / 0.364 & 23.19 / 0.315 & 23.42 / \blue{0.310} & 24.48 / 0.336 & \red{\textbf{25.07}} / \red{\textbf{0.293}}\\

\hline \hline
\multicolumn{9}{c}{Image Denoising \cite{zhang2017beyond}}\\
\hline
DIV2K &29.75 / 0.233 & 29.55 / 0.236& 29.47 / 0.275 & 29.62 / 0.263 & \blue{30.80} / \blue{0.215} & 29.61 / \blue{0.215} & 30.05 / 0.225& \red{\textbf{31.25}} / \red{\textbf{0.192}}\\
DPED   &30.24 / 0.218 & 29.87 / 0.230& 29.48 / 0.261 & 29.60 / 0.255 & \blue{31.23} / 0.195 & 30.09 / \blue{0.191} & 31.15 / 0.203& \red{\textbf{31.36}} / \red{\textbf{0.181}}\\
BSD &  29.92 / 0.240& 29.71 / 0.248 & 29.39 / 0.285 & 29.55 / 0.262 & \blue{30.40} / \blue{0.203} & 29.81 / \blue{0.203} & 30.39 / 0.214 & \red{\textbf{30.42}} / \red{\textbf{0.192}} \\

\hline \hline
\multicolumn{9}{c}{JPEG Artefact Removal \cite{zhang2017beyond}}\\
\hline
DIV2K   & 26.50 / 0.303 & \blue{26.71} / 0.295 & 26.32 / 0.295 & 26.37 / 0.301 & 26.48 / 0.315 & 26.27 / \blue{0.281} & 26.50 / 0.299 & \red{\textbf{26.77}} / \red{\textbf{0.261}}\\
DPED   & 26.20 / 0.305 & \blue{26.15} / 0.301 & 25.95 / 0.298 & 26.05 / 0.305 & 26.01 / 0.307 & 25.87 / \blue{0.296} & 26.12 / 0.305 & \red{\textbf{26.53}} / \red{\textbf{0.276}}\\
BSD &25.64 / 0.316 & \blue{25.71} / 0.310 & 25.43 / 0.309 & 25.49 / 0.313 & 25.54 / \blue{0.308} & 25.39 / \blue{0.308} & 25.52 / 0.312 & \red{\textbf{25.75}} / \red{\textbf{0.293}}\\


\hline
\end{tabular}
}
\end{table*}

\begin{figure}[!th]
    \centering
   {\includegraphics[trim={0cm 0cm 1.25cm 1.05cm}, clip, width=0.995\linewidth]{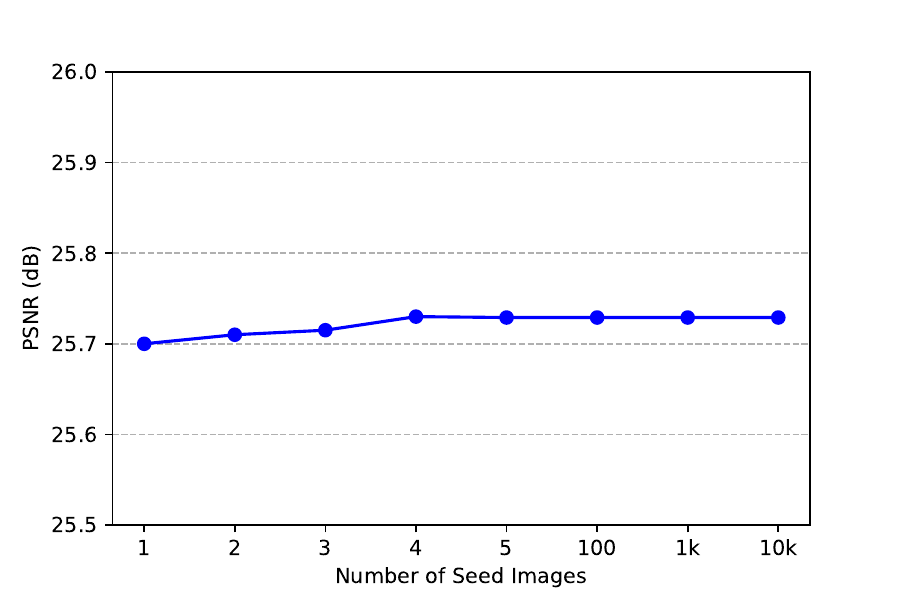} }
    \caption{Performance of EDSR model with the increasing the number of seed images used for training the MDF loss function. Note that PSNR increases only by 0.03\,dB and saturates for larger number of images. The inference results are reported for the BSD dataset.}
    \label{fig:number_of_seed_images}
\end{figure}

\paragraph{Number of seed images}
Next we investigate the impact of increasing the number of seed images while training the MDF loss function. The plot in Fig.~\ref{fig:number_of_seed_images} shows that the performance of EDSR increases by only 0.03\,dB when trained on 4 images and then it saturates. We did not observe any improvement in visual quality. Because the increase in performance in negligible when adding more seed images, we used a single image for training in our results. 


\begin{figure*}[t]
    \centering
  {\includegraphics[trim={0cm 10.8cm 12.7cm 0cm}, clip, width=0.995\linewidth]{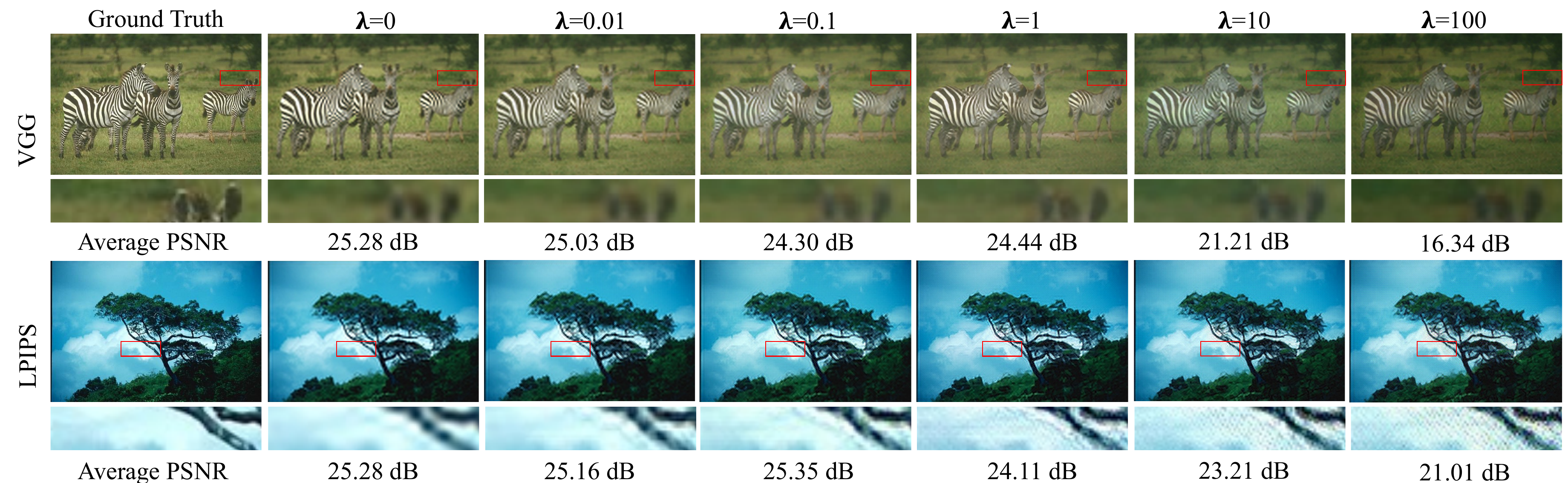} }
    \caption{Comparison of the \ac{SISR} results (EDSR) when trained using a weighted sum of VGG/LPIPS and MSE feature-wise losses: $\textrm{MSE} + \lambda\,\textrm{VGG/LPIPS}$. The average PSNR is reported for the entire test set.}
    \label{fig:ablation-alpha}
\end{figure*}

\section{Hyper-parameter tuning for VGG and LPIPS}
In Fig. \ref{fig:ablation-alpha} we show the qualitative results for the trade-off between the MSE and LPIPS/VGG network components in the joint loss function. For fair comparison, we conducted a hyper-parameter search over the scalar $\lambda$ controlling the weight of the feature-wise loss function. We searched over the values in \{$\lambda: \lambda=10^{k},\,k=-3,..,3$\}. The greater $\lambda$ parameter is, the more LPIPS/VGG components contribution is. In our experiments across all image restoration applications, we found the best results are produced when $\lambda = 1$ for VGG and $\lambda = 0.1$ for LPIPS loss. Additional qualitative results are provided in the HTML report.

\section{Image quality metrics and loss functions}
\begin{figure*}[th]
    \centering
   {\includegraphics[ width=.9\linewidth]{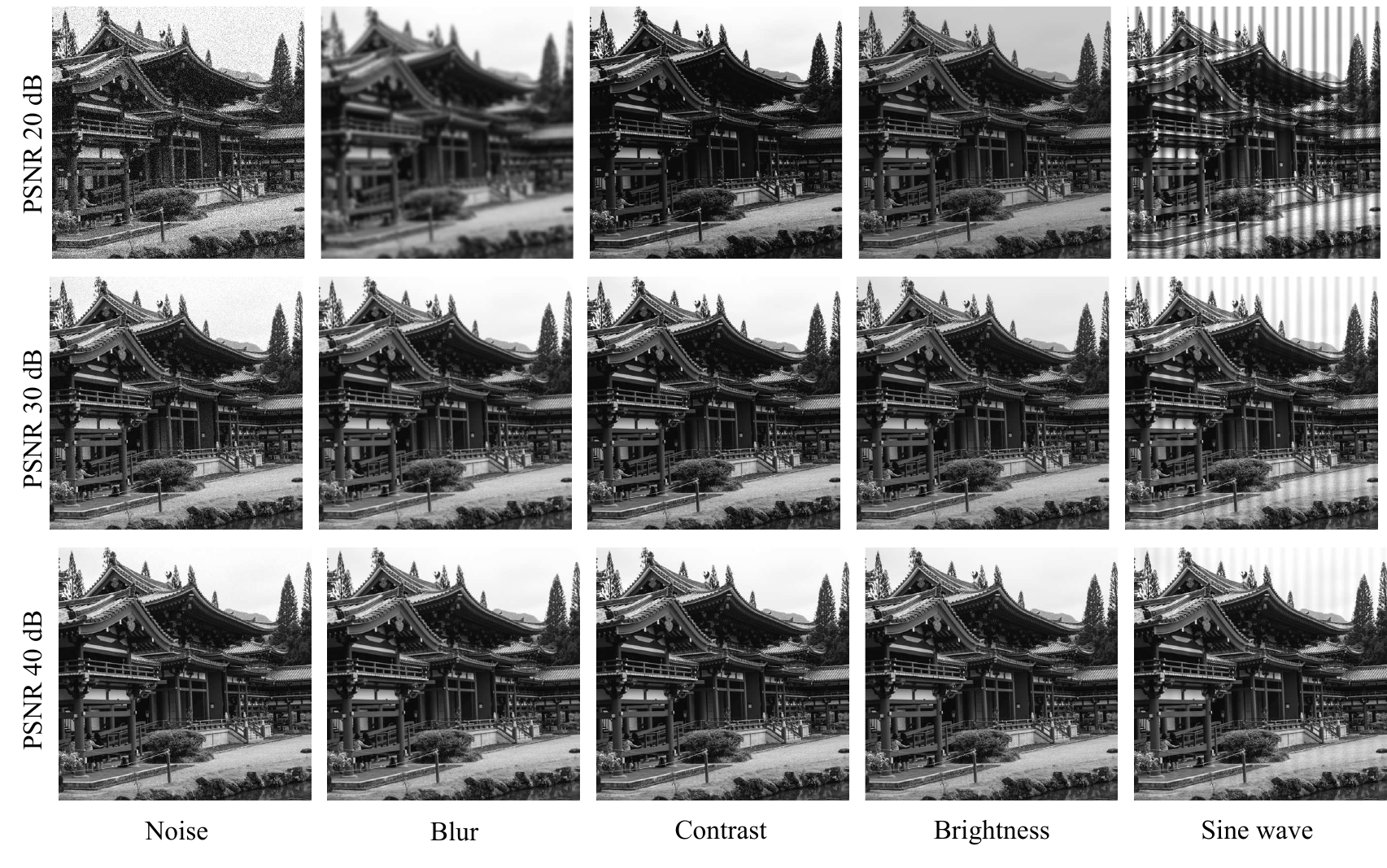} }
    \caption{Examples of images used to test the sensitivity of  loss functions to different types of distortions. We introduced artifacts so that the each distortion results in the same PSNR level (across each row). Here we provide examples of images at 20\,dB, 30\,dB and 40\,dB. Note that the perceived quality differs between the columns despite the same PSNR level.}
    \label{fig:distortion}
\end{figure*}

To further investigate the performance of loss functions as quality predictors, we generated a set of images that were distorted by blur, noise, added sinusoidal grating, contrast and brightness changes. The distortions were generated so that they degraded the image in equal steps of PSNR. Fig.~\ref{fig:distortion} presents an example of images with introduced distortions at three PSNR levels. The experiment shows a failure case of PSNR, predicting the same quality even though the distortions due to contrast and  brightness are much less objectionable than the others to a human observer.

\begin{figure*}[t]
    \centering
   {\includegraphics[trim={0cm 0cm 0cm 0cm}, clip, width=0.995\linewidth]{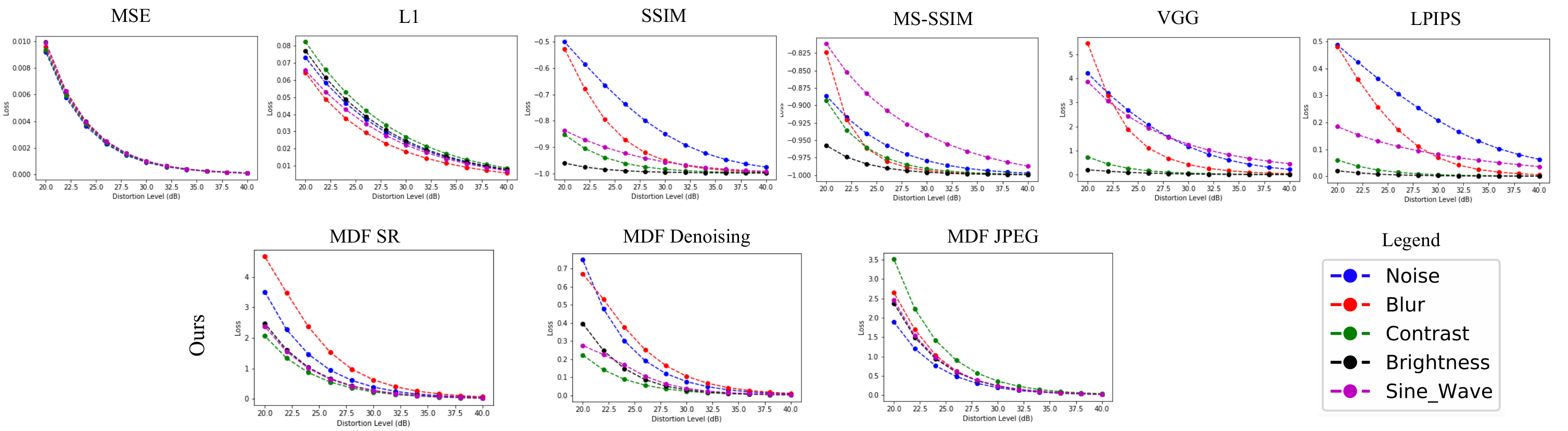} }
    \caption{Loss values for the increasing amount of distortions of different types. The distortion levels have been generated to result in equal PSNR values, shown on the x-axis. Despite the same PSNR value, the distortions due to noise, blur and added sinusoidal wave are much more noticeable than those due to contrast and brightness change (refer to Fig.~\ref{fig:distortion}). The MDF loss accurately predicts the perceived magnitude of task specific distortions for which it is trained.
    }
    \label{fig:artificial-dist-quality}
\end{figure*}

In Fig.~\ref{fig:artificial-dist-quality}, we show the loss values computed for the increasing amount of distortions of different types for different loss functions. Despite the same PSNR value, the distortions due to noise, blur and added sinusoidal wave are much more noticeable than those due to contrast and brightness change (refer to Fig.~\ref{fig:distortion}). The loss functions derived from quality metrics (SSIM, MS-SSIM) and also feature-wise losses (VGG, LPIPS) penalize more the distortions that result in higher degradation of quality. In contrast, MDF losses penalize the most the distortions that are relevant for a given task: blur in case of SISR (MDF~SR), blur and noise in case of denoising, and contrast followed by the mixture of all distortions in case of JPEG artifact removal. This is another example demonstrating that an effective loss (MDF) function does not need to predict image quality. 

\section*{Acknowledgements}
This project has received funding from the European Research Council (ERC) under the European Union’s Horizon 2020 research and innovation programme (grant agreement N$^\circ$ 725253–EyeCode).

\clearpage
{\small
\bibliographystyle{ieee_fullname}
\bibliography{egbib}
}

\end{document}